# TOWARD IMPROVING PREDICTION OF SEDIMENT TRANSPORT OVER WAVE-INDUCED RIPPLES

Rafik ABSI[1], Hitoshi TANAKA[2]

**ABSTRACT**

Sediment transport over wave-induced ripples is a very complex phenomenon where available models fail to provide accurate predictions. For coastal engineering applications, the 1-DV advection-diffusion equation could be used with an additional parameter α related to the process of vortex shedding above ripples (Absi, 2010). The aim of this study is to provide simple practical analytical tools. An analytical eddy viscosity profile was validated by DNS data of turbulent channel flows (Absi *et al.*, 2011). In this study, we will show that: (1) the period-averaged eddy viscosity in oscillatory boundary layers could be described by this simple analytical formulation; (2) The shape of the vertical profile is validated by period-averaged eddy viscosity of baseline (BSL) k-ω model (Suntoyo and Tanaka, 2009) for sinusoidal and asymmetric waves; (3) The vertical eddy viscosity profile depends on the wave non-linearity parameter and requires therefore a specific calibration.

## 1. INTRODUCTION

Coastal zones are of high vulnerability to natural hazards/disasters. Waves, currents and tides make coastal zones areas changing due to erosion and deposition of sediments. Thus, an understanding of sediment transport process in coastal zones is of crucial importance for accurate predictions of coast line evolution and sea-bed changes. However, the modeling of coastal sediment transport needs a compromise between two types of models: detailed mathematical models and engineering approaches. This compromise is imposed by on the one hand the accuracy of predictions and on the other hand the usability in practical applications (Absi 2011). In coastal engineering practical accurate engineering models which take into account the more important involved physics, are needed.

In the engineering approach, the net (averaged over the wave period) total sediment transport is obtained as the sum of the net bed load and net suspended load transport rates (Fredsoe and Deigaard 1992). For suspended load, the net sand transport is defined as the sum of the net current-related and the net wave-related transport components. The wave-related suspended transport component requires computation of the time-averaged suspended sediment concentration (SSC) profile and its integration in the vertical direction (van Rijn 2007). Computation of SSC needs the sediment diffusivity $\varepsilon_s$ which is related to the eddy viscosity $\nu_t$ by the parameter β (i.e., the inverse of the turbulent Schmidt number).

For moderate wave conditions and/or deep water, wave ripples can be formed on the sea bottom. If the ripples are relatively steep ($\eta_r/\lambda_r \geq 0.12$, where $\eta_r$ is the ripple height and $\lambda_r$ is the ripple wavelength), the mixing close to the bed is dominated by coherent, periodic vortex structures. Above rippled beds, the mixing in the near bed layer is dominated by the mechanism of vortex shedding which entrains sediments.

The aim of our study is to improve the prediction of SSC over ripples by using simple analytical tools which take into account the more important involved physics, for practical use in coastal engineering.

## 2. TIME-AVERAGED CONCENTRATIONS OVER WAVE-INDUCED RIPPLES

Sediment diffusivity $\varepsilon_s$ describes the disorganized ''diffusive'' process. The process of vortex formation and shedding at flow reversal above ripples is a relatively coherent phenomenon. The associated convective sediment entrainment process may also be characterized as coherent, instead of a pure disorganized ''diffusive'' process represented in the classical gradient diffusion model (Thorne *et al.* 2002). Nielsen (1992) indicated that both convective and diffusive mechanisms are involved in the entrainment processes. In the combined convection-diffusion formulation, the steady state advection-diffusion equation is given by

$$w_s c + \varepsilon_s \frac{dc}{dz} + F_{conv} = 0 \qquad (1)$$

[1] **EBI, Inst. Polytech. St-Louis, Cergy University, 32 Bd du Port, 95094 Cergy-Pontoise France r.absi@ebi-edu.com**
[2] **Department Civil Eng., Tohoku University, 6-6-06 Aoba, Sendai 980-8579, Japan tanaka@tsunami2.civil.tohoku.ac.jp**

The respective terms in (1) represent downward settling, upward diffusion (given by gradient diffusion $F_{diff} = \varepsilon_s (dc/dz)$) and upward convection $F_{conv}$. The upward convection term $F_{conv}$ was given by Thorne *et al.* (2002) as $F_{conv} = -w_s c_0 F(z)$, where F(z) is a function describing the probability of a particle reaching height z above the bed (Nielsen 1992). Thorne *et al.* (2009) wrote $F_{conv} = -\overline{w_w c_w}$ where $w_w$ and $c_w$ are periodic components respectively of concentrations and vertical velocity and the overbar denotes time averaging. It is possible to write (1) in the form of a diffusion equation. The time-averaged (over the wave period) advection-diffusion equation is given therefore by (Absi 2010)

$$w_s c + \varepsilon_s^* \frac{dc}{dz} = 0 \qquad (2)$$

where $\varepsilon_s^* = \alpha \varepsilon_s$ and $\alpha$ is a parameter related to convective sediment entrainment process associated to the process of vortex shedding above ripples $\alpha = 1/(1 + F_{conv}/(w_s c))$. With the upward convection $F_{conv} = -w_s c_0 F(z)$, $\alpha$ becomes equal to $1/(1 - (c_0/c)F(z))$, while with $F_{conv} = -\overline{w_w c_w}$ (Thorne *et al.*, 2009) $\alpha = 1/(1 - \overline{w_w c_w}/(w_s c))$. The condition of Sheng and Hay (1995) $\overline{w_w c_w}/(w_s c) < 0.2$ shows therefore that when the convective transfer is very small (above low steepness ripples), $\alpha \approx 1$ and therefore $\varepsilon_s^* \approx \varepsilon_s$ (Absi, 2011). From equations (1) and (2), it is possible to write

$$\varepsilon_s^* = \left(1 + \frac{F_{conv}}{F_{diff}}\right) \varepsilon_s \qquad (3)$$

and therefore $\alpha = 1 + (F_{conv}/F_{diff})$. This equation shows that $\alpha$ depends on the relative importance of coherent vortex shedding (related to $F_{conv}$) and random turbulence (related to $F_{diff}$). When $F_{conv} > F_{diff}$ $=> \alpha > 1$, while $F_{conv} << F_{diff}$ $=> \alpha \approx 1$ and therefore $\varepsilon_s^* \approx \varepsilon_s$. Absi (2010) proposed the following equation

$$\frac{d^2 \ln c}{dz^2} = \frac{w_s}{\varepsilon_s^{*2}} \frac{d\varepsilon_s^*}{dz} \qquad (4)$$

Eq. (4) provides a link between upward concavity/convexity of concentration profiles (in semi-log plots) and increasing/decreasing of $\varepsilon_s^*$. Increasing $\varepsilon_s^*$ allows upward concave concentration profile, while decreasing $\varepsilon_s^*$ allows an upward convex concentration profile.

In order to allow adequate predictions of suspended sediment transport, it is important to understand interaction between sediment particles and turbulence of fluid flow. The turbulent diffusion of suspended sediments $\varepsilon_s$ is given by

$$\varepsilon_s = \beta \nu_t \qquad (5)$$

where $\beta$ = inverse of the turbulent Schmidt number, describes the difference between diffusivity of momentum (diffusion of a fluid "particle") and diffusivity of sediment particles. It should depend on the particle Stokes number (Absi et al. 2011). However, for simplicity and in order to allow analytical analysis, we suggested a simple equation $\beta = \beta_b \exp(C_\beta z/\delta)$; where $\beta_b$ = the value of $\beta$ close to the bed and $C_\beta$

= coefficient (Absi 2010). This $\beta$ (y) profile increases with z for $C_\beta > 0$ and decreases for $C_\beta < 0$. We used an analytical eddy viscosity given by

$$\nu_t = \kappa u_* z\, e^{-C \frac{z}{\delta}} \quad (6)$$

where $u_*$ = the friction velocity (m/s), $\delta$ = the boundary layer thickness (m), $\kappa$ = the Karman constant (=0.41) and C a parameter =1.12 (Hsu and Jan 1998, Absi 2000, Absi 2010). Using the $\beta$-function and eddy viscosity (6), the sediment diffusivity is given therefore by

$$\varepsilon_s = A_s z\, e^{-\frac{z}{B_s}} \quad (7)$$

where $A_s = \kappa \beta_b u_*$ (m/s) and $B_s = \delta/(C - C_\beta)$ (m). Absi (2010) suggested an empirical function for $\alpha$ given by $\alpha = 1 + D \exp(-z/h_s)$; where D and $h_s$ are two parameters.

**Test case: Fine and coarse sediments over rippled beds in the same flow (McFetridge and Nielsen, 1985)**

Maximum value of the free stream velocity, wave period, mean depth of the flow, orbital amplitude or near-bed flow semi-excursion, mean ripple height, mean ripple wavelength, equivalent roughness $k_s = 25\,\eta_r\,(\eta_r/\lambda_r)$, friction factor $f_w = 0.237\,(k_s/a_m)^{0.52}$ (Soulsby, 1997), mean magnitude of the friction velocity in the wave cycle $u_* = 0.763\,(f_w/2)^{0.5}\,U_0$ (Davies, 1986) and $\delta = u_*/\sigma$ are given respectively in table 1.

**Table 1.** Flow parameters

| $U_0$(cm/s) | $T$(s) | $h$(m) | $a_m$(cm) | $\eta_r$(cm) | $\lambda_r$(cm) | $k_s$(cm) | $f_w$ | $u_*$(cm/s) | $\delta$(cm) |
|---|---|---|---|---|---|---|---|---|---|
| 27.8 | 1.51 | 0.3 | 6.68 | 1.1 | 7.8 | 3.88 | 0.178 | 6.3 | 1.5 |

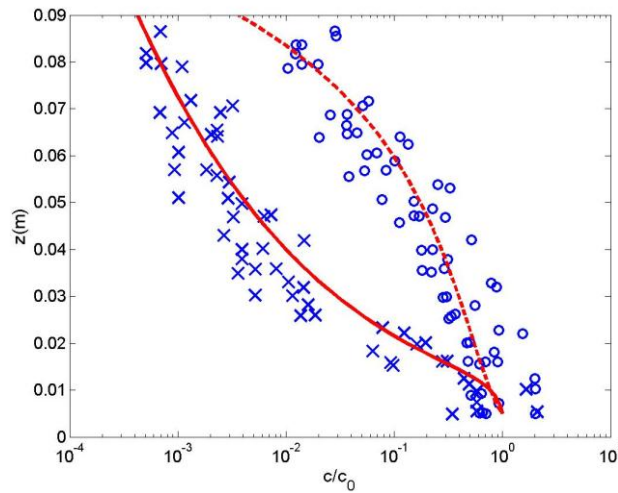

**Figure 1:** *Time-averaged concentration profiles over wave-induced ripples. Symbols: measurements (McFetridge and Nielsen, 1985), (○) fine; (×) coarse; Curves: solutions of Eq. (2) (Absi 2010).*

Figure 1 shows time-averaged concentrations of fine (○) and coarse (×) sediments suspended by waves over ripples. The present method allows a good description of concentration profiles for both fine (dashed line) and coarse sand (solid line). The parameters were chosen to give a good fit.

For an eddy viscosity given by: $\nu_t = 0.0258 \times z \times \exp(-1.12 \times z/0.015)$, the parameters are for fine sediments: $\beta_b = 0.97 \approx 1$, $C_\beta = 0.438$ and therefore $\varepsilon_s = 0.025 \times z \times \exp(-z/0.022)$ and for coarse sediments: $\beta_b = 0.659$, $C_\beta = 1.1$ and therefore $\varepsilon_s = 0.017 \times z \times \exp(-z/0.75)$ with $\alpha = 1 + 403 \exp(-z/0.002)$. The value $C_\beta = 0.438$ for fine sediments could be related to an inaccurate estimation of the boundary layer thickness. For coarse sediments, the profile of $\varepsilon_s^*$ (Absi 2010, solid line in figure 7) shows the effect of parameter $\alpha$ which indicates that vortex shedding occurs at z <0.015m.

However in order to allow practical use for predictive purpose, the method needs calibration. Before calibrating parameters of α and β, we need to assess and validate the eddy viscosity profile given by Eq. (6).

## 3. ANALYTICAL EDDY VISCOSITY FORMULATION

Eq. (6) was used as an empirical equation. However in order to allow more general use, we need a deeper theoretical analysis.

**Steady plan channel flows: analysis by DNS data**

In the equilibrium region z+>50, the turbulent kinetic energy (TKE) is given by $k \approx u_* \exp(-C_k z/\delta)$ (Nezu and Nakagawa 1993). Since in the inner region the streamwise velocity profile is given by the log-law, it is possible to write a mixing length as $l_m = \kappa z \exp(-C_k z/\delta)$ and therefore Eq. (6) for eddy viscosity. Figure 2 shows TKE profiles given by two analytical solutions (Nezu and Nakagawa 1993, Absi 2008) and eddy viscosity profiles (white dashed lines) given by Eq. (6). In figure 2, variables with the superscript of + are those nondimensionalized by the friction velocity and the kinematic viscosity as $z^+ = z u_*/\nu$; $k^+ = k/u_*$; $\nu_t^+ = \nu_t/\nu$. Comparisons with DNS data (data of Iwamoto 2002, Iwamoto et al. 2002, Hoyas and Jiménez 2006) show that Eq. (6) provides accurate description of DNS in the equilibrium region (Absi et al. 2011).

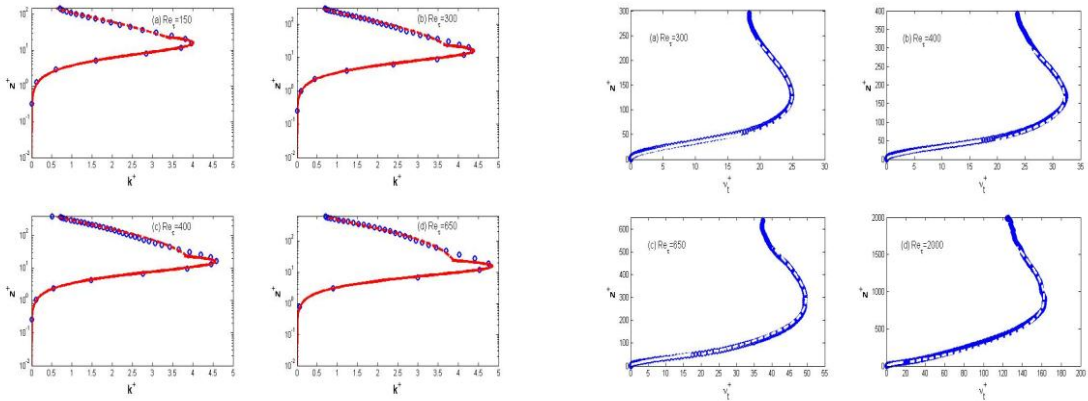

**Figure 2:** *Turbulent kinetic energy (left) and eddy viscosity (right) profiles in plan channel flows for different friction Reynolds numbers. Symbols: DNS data; Lines: analytical (Absi, 2008; Absi et al., 2011).*

**Oscillatory flows: analysis by a two-equation model**

Eq. (6) for eddy viscosity was validated for the case of steady plane channel flow. However for use in wave boundary layers, we need to assess this equation for the case of oscillatory flows. Eq. (6) is therefore analyzed by the baseline (BSL) k-ω model. This model allows accurate prediction of velocity profiles in oscillatory boundary layers (Suntoyo and Tanaka 2009).

Figure (3.a) presents temporal and spatial variation of dimensionless eddy viscosity for a sinusoidal wave. Figure (3.b) shows comparison between period-averaged eddy viscosity obtained by BSL k-ω model

(symbols) and analytical profile of Eq. (6) (dashed line). Even if the eddy viscosity is highly time-dependent (figure 3.a), the period-averaged dimensionless eddy viscosity (Figure 3.b) has a shape which is well described by the analytical profile given by Eq. (6) for $z/z_h < 0.6$ (figure3.b) where $z_h$ is the water depth or the distance from the wall to the axis of symmetry or free surface.

Figure (3.c) presents temporal and spatial variation of dimensionless eddy viscosity for asymmetric waves. Figure (3.d) shows comparison between period-averaged eddy viscosity obtained by BSL k-ω model (symbols) and analytical profile of Eq. (6) (dashed line). Even for the case of asymmetric wave, the period-averaged dimensionless eddy viscosity has a shape which is well described by Eq. (6) for $z/z_h < 0.5$ (figure3.d).

Figures (3.b) and (3.d) shows that the period-averaged eddy viscosity profile for sinusoidal wave is different from the profile for asymmetric wave. This indicates that the period-averaged eddy viscosity profile should depend on the wave non-linearity parameter given by $N_i=U_c/\hat{u}$, where $U_c$ is the velocity at wave crest and $\hat{u}$ is the total velocity amplitude. We need therefore a specific calibration for parameters of Eq. (6) using full-range equations of friction coefficient (Tanaka and Thu 1994) and wave boundary layer thickness (Sana and Tanaka 2007).

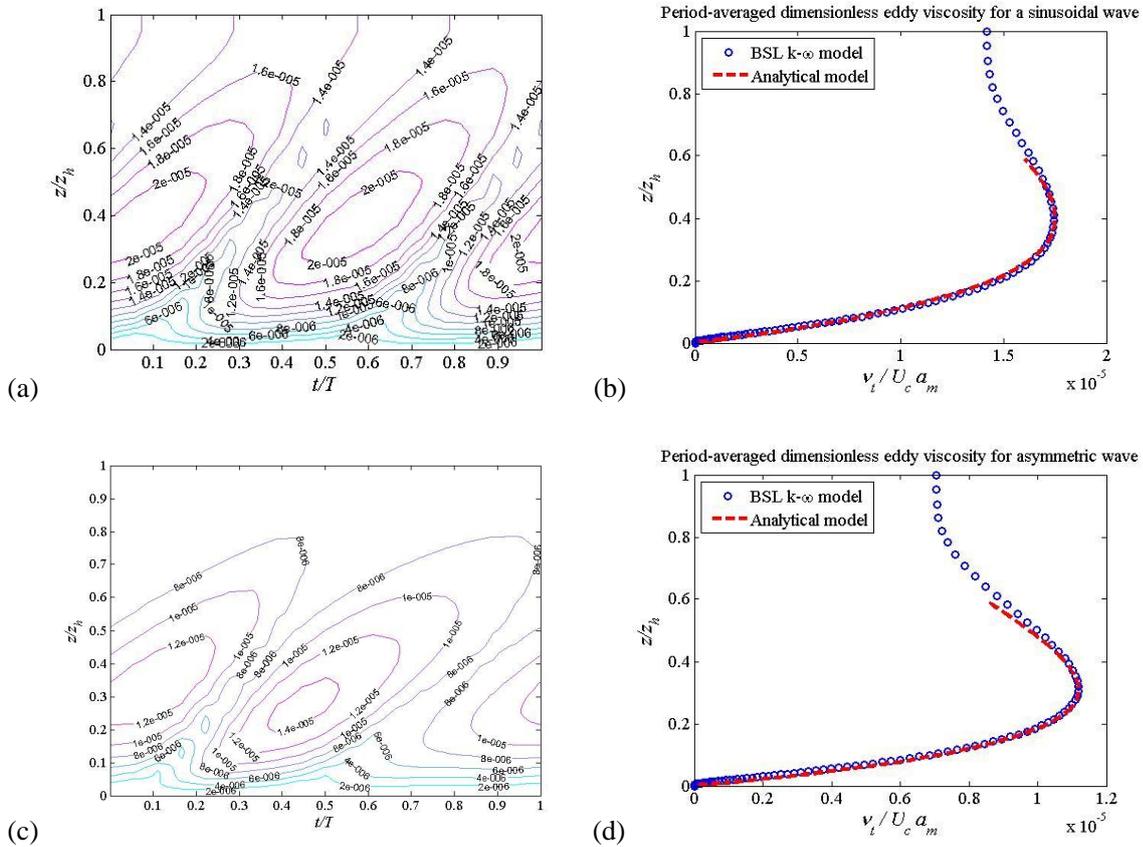

**Figure 3:** *Dimensionless eddy viscosity; Left: Temporal and Spatial Variation; Right: Period-averaged dimensionless eddy viscosity; Top: sinusoidal wave; Bottom: asymmetric wave.*

### 3. CONCLUSIONS

The main conclusions of the present study are:
- A modified advection-diffusion equation with an additional parameter α related to the process of vortex shedding above ripples allows a good description of suspended sediment concentration profiles
- For practical applications the period-averaged eddy viscosity could be described by a simple analytical formulation

- The shape of the analytical period-averaged eddy viscosity formulation was validated by BSL k-ω model for sinusoidal and asymmetric waves
- Period-averaged eddy viscosity profile depends on the wave non-linearity parameter and requires therefore a specific calibration.

**ACKNOWLEDGMENTS**

The first author is grateful for the financial support provided by Japan Society for the Promotion of Science (JSPS), within the FY2010 JSPS Invitation Fellowship Program for Research in Japan (No. S-10168).